\documentclass[final,5p,times,twocolumn]{elsarticle}

 \biboptions{sort&compress}

\usepackage{inputenc}
\usepackage{verbatim}
\usepackage{epsfig}
\usepackage{bm}
\usepackage{xcolor}
\usepackage{float}
\usepackage{dcolumn} 
\usepackage{physics}
\usepackage{graphicx}
\usepackage{capt-of}
\usepackage{amsmath}
\usepackage{amssymb}
\usepackage{relsize}
\usepackage[colorlinks,breaklinks]{hyperref}
\usepackage{xcolor}

\usepackage{multirow}

\usepackage{comment}

\newcommand{\be}{\begin{equation}}
\newcommand{\ee}{\end{equation}}
\newcommand{\bea}{\begin{eqnarray}}
\newcommand{\eea}{\end{eqnarray}}

\begin{document}

\begin{frontmatter}

\title{Data-driven analysis of dipole strength functions using artificial neural networks}

\author[kph]{Weiguang Jiang}
\ead{wjiang@uni-mainz.de}
\author[kph]{Tim Egert}
\ead{tiegert@students.uni-mainz.de}
\author[kph,him]{Sonia Bacca}
\ead{s.bacca@uni-mainz.de}
\author[kph,frib,ornl]{Francesca Bonaiti}
\ead{bonaiti@frib.msu.edu}
\author[da]{Peter von Neumann Cosel}
\ead{vnc@ikp.tu-darmstadt.de}
\address[kph]{Institut f\"{u}r Kernphysik and PRISMA$^+$ Cluster of Excellence, Johannes Gutenberg-Universit\"at, 55128 Mainz, Germany}
\address[him]{Helmholtz-Institut Mainz, Johannes Gutenberg-Universit\"at Mainz, D-55099 Mainz, Germany}
\address[frib]{Facility for Rare Isotope Beams, Michigan State University, East Lansing, MI 48824, USA}
\address[ornl]{Physics Division, Oak Ridge National Laboratory, Oak Ridge, TN 37831, USA}
\address[da]{Institut f\"{u}r Kernphysik, Technische Universität Darmstadt, D-64289 Darmstadt, Germany}

%%%%%%%%%%%%%%%%%%%%%%%%%%%%%%%%%%%%%%%%%%%%%%%%%%%
%%%%%%%%%%%%%%%%%%%%%%%%%%%%%%%%%%%%%%%%%%%%%%%%%%%

\begin{abstract}
We present a data-driven analysis of dipole strength functions across the nuclear chart, employing an artificial neural network to model and predict nuclear dipole responses. We train the network on a dataset of experimentally measured dipole strength functions for 216 different nuclei. To assess its predictive capability, we test the trained model on an additional set of 10 new nuclei, where experimental data exist. Our results demonstrate that the artificial neural network not only accurately reproduces known data but also identifies potential inconsistencies in certain experimental datasets, indicating which results may warrant further review or possible rejection. Additionally, for nuclei where experimental data are sparse or unavailable, the network confirms theoretical calculations, reinforcing its utility as a predictive tool in nuclear physics. Finally, utilizing the predicted electric dipole polarizability, we extract the value of the symmetry energy at saturation density and find it consistent with results from the literature.
\end{abstract}

%%%%%%%%%%%%%%%%%%%%%%%%%%%%%%%%%%
%%%%%%%%%%%%%%%%%%%%%%%%%%%%%%%%%%


\end{frontmatter}

%%%%%%%%%%%%%%%%%%%%%%%%%%%%%%%%%%%%%%%%%%%%%%%
% Introduction
%%%%%%%%%%%%%%%%%%%%%%%%%%%%%%%%%%%%%%%%%%%%%%%
\section{Introduction}

Dipole strength functions play a fundamental role in nuclear physics, providing insight into the dynamics of the nucleus~\cite{BaccaPastore}. These functions describe the probability distribution of dipole transitions and are crucial not only to nuclear structure and reaction dynamics, but are also important for a wide range of fields such as nuclear astrophysics \cite{goriely23}, medical isotope production \cite{wang22}, fission \cite{chadwick11} and fusion reactor \cite{meschini23} technologies and waste transmutation \cite{salvatore11}.
One of the most significant applications in astrophysics lies in the study of the formation of heavy elements in stellar environments. Dipole responses, especially the pygmy dipole resonances at low energy~\cite{Savran:2013bha}, contribute to the neutron capture rates, influencing nucleosynthesis pathways in explosive astrophysical events~\cite{Arnould_2007}.

An integral quantity related to the dipole strength function, defined as the inverse energy weighted sum rule of the dipole response, is the electric dipole polarizability.  
This observable is correlated to the neutron skin thickness \cite{tamii11}---the difference between the radii of the neutron and proton distributions---and serves as a probe of the symmetry energy \cite{rocamaza13}, a crucial component of the nuclear equation of state (EoS). Understanding the symmetry energy is particularly important in astrophysics, both for the structure of neutron stars~\cite{lat21} and the dynamics of supernovae~\cite{yas20}.

Experimental data for dipole strength functions, and consequently of the electric dipole polarizability, are limited, particularly for exotic nuclei far from stability. 
Close to the line of stability, dipole strength functions can be extracted from measurements of photoabsorption cross sections~\cite{EXFOR} and other electromagnetic processes, such as relativistic Coulomb excitation in forward-angle $(p,p')$ reactions~\cite{vnc19a}. 
Furthermore, the low-energy strength can be accessed via nuclear resonance fluorescence (NRF) experiments \cite{zilges22} and indirect techniques such as the Oslo method~\cite{larsen19}.
Recently, accurate determinations of the electric dipole polarizability have also been provided, primarily from $(p,p')$ reactions, however only for some selected nuclei~\cite{tamii11,bassauer20,martin17,brandherm24,bir17,fea23}. 
These experiments have driven the development of new theoretical approaches, which may also be used where data do not exist.
Presently, the situation is such that modern calculations either based on {\it ab initio} nuclear theory~\cite{hag16,miorelli2016,miorelli2018, bonaiti2022, bonaiti2024} or density functional theory \cite{roc18} (and references therein) are available, but can not always be tested against experimental data.
While awaiting new experiments to be carried out, an alternative path can be pursued.

Recent advances in artificial intelligence (AI), specifically artificial neural networks (ANNs), offer new opportunities for predicting nuclear properties from available experimental data \cite{utama2016,niu2018,wang2019,lasseri2020}. ANNs can be trained on existing datasets to model complex correlations in nuclear data, providing an efficient tool for extrapolating dipole strength functions or polarizabilities to unmeasured nuclei. 
With ANNs, one is also able to estimate confidence intervals on predictions \cite{Jiang2019,DONG2023,Sobczyk2024}. 
By modeling uncertainties, ANNs can highlight areas where predictions may be less certain, guiding future experiments and enabling more robust theoretical validation across diverse datasets.

This letter presents the first data-driven analysis of the dipole strength across the nuclear chart using ANNs trained on available experimental data. 
We train and validate the network on a set of 216 nuclei, and then test it on an additional 10, emphasizing how ANN predictions compare against existing experiments and recent theoretical calculations. To connect to nuclear matter, we constrain the symmetry energy by fitting our ANN predictions of the electric dipole polarizability to a phenomenological model.

\section{Strategy}

Preparing the training dataset for the neural network is critical to achieve reliable predictions for the targeted observables.
In this regard, a few preliminary considerations need to be made. 
The dipole polarizability, typically denoted with $\alpha_D$, is defined as
\begin{eqnarray}
\label{alphaD}
 \alpha_D = \frac{3(\hbar c)^3}{2} \int_{0}^{\infty} \frac{f_{E1}(E)}{E}d E \,,
\end{eqnarray}
where $f_{E1}$ is the dipole strength function depending on the energy $E$, which runs up to infinity. We are in principle interested in both $f_{E1}$ and $\alpha_D$. However, if the objective were only $\alpha_D$, one could think of training an ANN directly on $\alpha_D$ data, which however turns out to be difficult due to the following reasons.
The first major obstacle is the scarcity of available experimental data for training. The existing experimental measurements of $\alpha_D$ are limited to only about 150 data points. Such a limited dataset raises challenges for the training of any machine learning model, as it most likely leads to overfitting, where the model memorizes the training data rather than learning to generalize.
The second challenge arises from the cumulative nature of the observable. Equation~(\ref{alphaD}) makes $\alpha_D$ particularly sensitive to the energy range over which the integration is performed. Experimental data occasionally have discrepancies and inconsistencies in the measured values of $\alpha_D$, depending on the range of integration. Such inconsistencies complicate the direct use of $\alpha_D$ as a training target of an ANN, as the model could inadvertently learn the biases introduced by the various energy ranges rather than the underlying physical trends.

We focus therefore on the dipole strength function $f_{E1}$, rather than on $\alpha_D$, since the latter can then be computed from Eq.~(\ref{alphaD}) from any given $f_{E1}$. In the case of $f_{E1}$, each data set from a given experiment offers as many data as the available points in $E$, which quickly turns into an increase of a couple of orders of magnitude in the number of data points with respect to the $\alpha_D$ dataset. 
Furthermore, $f_{E1}$ has been investigated in more nuclei, albeit perhaps only within selected energy regions, than the number of nuclei for which we have data on $\alpha_D$. Since ANNs typically perform better with larger datasets as more data provides it with richer patterns to learn from, this is clearly the best choice.

\subsection{Dataset preparation }
Having decided to use $f_{E1}$ as the input data for training our ANNs, we need to prepare the dataset.
There exist several data where one could extract $f_{E1}(E)$ from. The largest portion of them comes from measurements of the photoabsorption or photo-neutron cross sections $\sigma_{\gamma}(E)$ at different gamma-ray energies $E$. Under the assumption that the process occurs primarily via an electric dipole operator---valid below about 100 MeV of energy \cite{BaccaPastore} ---the dipole strength function can be extracted as 
\begin{eqnarray}
\label{f1}
f_{E1}(E) = \frac{\sigma_{\gamma}(E)}{3(\pi \hbar c)^2 E}\,.
\end{eqnarray}
We take these data from the well-established EXFOR database~\cite{EXFOR} described in \cite{otuka14}.
Since the majority of experiments provide information above the lowest particle threshold only, we add results of NRF and Oslo-type experiments from the evaluated IAEA database on photon strength functions~\cite{IAEA} (for details of the evaluation see~\cite{refOslo}).
Additionally, results from the ($p,p^\prime$) experiments discussed above are included. 
% No need to cite, the publications have been cited in the introduction.
Overall, we create a dataset including 226 different nuclei, most of them being close to the line of stability. Note that among these nuclei, 66 of them contain only Oslo or NRF data which cover a small low-energy range. In total, the dataset includes 32,958 data points, each corresponding to dipole strength values at different energies. 

To enhance the robustness of the training process and maximize the utility of the available data, we implemented data augmentation techniques. Spline interpolation is applied independently to the data from each experiment, without mixing or interpolating between different experiments. After augmentation each isotope is represented by 1000 data points, resulting in a dataset of 166,370 points across all nuclei. For rigorous model evaluation, the dataset was strategically divided into training, validation, and testing sets. In our approach, we choose 10 nuclei to serve as the testing set. These nuclei were intentionally excluded from any stage of the machine-learning process, ensuring that they were never exposed to the model during training or validation. The remaining data, comprising 216 nuclei, was then randomized (shuffled) and divided into two parts: $90\%$ was allocated to the training set, and $10\%$ was reserved for the validation set.

\subsection{Neural Network Architecture}
We employ a feedforward neural network with a fully connected architecture. The input layer includes the mass number $A$, proton number $Z$, and excitation energy $E$, with the output layer predicting the dipole strength function $f_{E1}$. Between the input and output layers, the network comprises four hidden layers with [16, 32, 32, 32] neurons, using the Rectified Linear Unit (ReLU) and Hyperbolic tangent (tanh) activation functions to capture non-linear patterns. The root mean squared error (RMSE) is used as the evaluation metric, aligning with the scale of the output variable and ensuring interpretable results.

The ANN weights are initialized using Glorot (Xavier) initialization to maintain variance across layers, and training is conducted with the adaptive moment estimation (Adam) optimizer. Note that the initial weights and the starting learning rate should be appropriately scaled relative to the size of the input data. If these parameters are not carefully chosen, there is a significant risk that the training process could become trapped in a local minimum or experience slow convergence. Early stopping with a patience of 50 epochs is employed to prevent overfitting by monitoring validation loss. More details about the ANNs and the diagnostic tools used can be found in the Supplementary Material.

\subsection{Uncertainty Quantification}
Once we have a functioning ANN, it is crucial to estimate the uncertainties embedded in the neural network predictions, as this helps us understand the limitations and reliability of our machine-learning model. The sources of error can generally be categorized into two main types: model uncertainty and data uncertainty. Model uncertainty arises from variability in the trainable parameters (also known as model discrepancy), influenced by differences in initialization and the stochastic nature of the training process, which can lead to varying outcomes even when trained on the same dataset. 

Data uncertainty stems from the quality and consistency of the input data, including inherent measurement errors and inconsistencies between experiments on the same nucleus. In this study, experimental errors were not included in the training due to the lack of error assignments for some data and the absence of correlation information between the errors.

To quantify such uncertainties, we employed ensemble learning \cite{opitz1999,polikar2006,sagi2018,ganaie2022}, which consists of training multiple neural network models with different initializations and randomized splits of the data between training and validation sets. The ANN prediction is taken as the median of the ensemble's predicted distribution, while the uncertainty is quantified by the 68$\%$ highest density interval \cite{Hyndman1996}, which may result in non-symmetric error bars. With this approach, we are able to account for parameter uncertainty and partially address data uncertainty by capturing the variability in the training data. Further details on the methods used for uncertainty quantification are provided in the Supplementary Material.

\section{Results}

\begin{figure}[h]
\includegraphics[width=1 \linewidth]{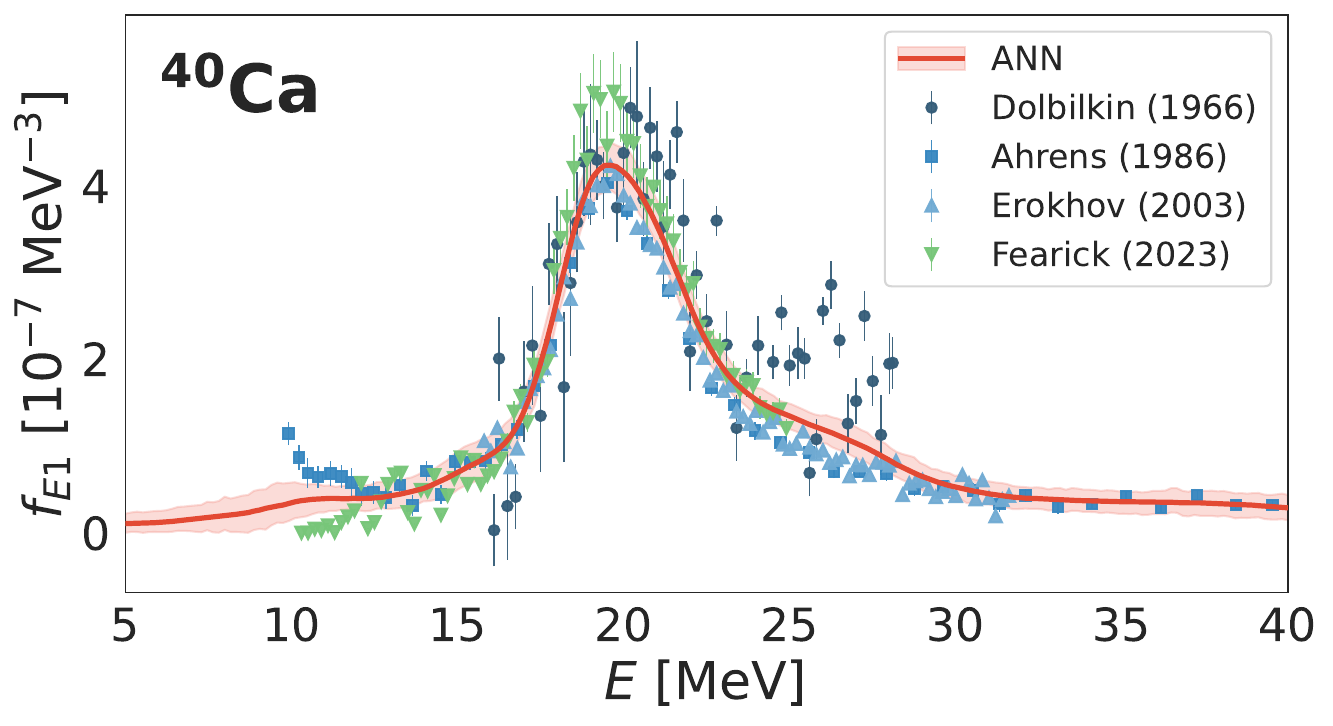}
\includegraphics[width=1 \linewidth]{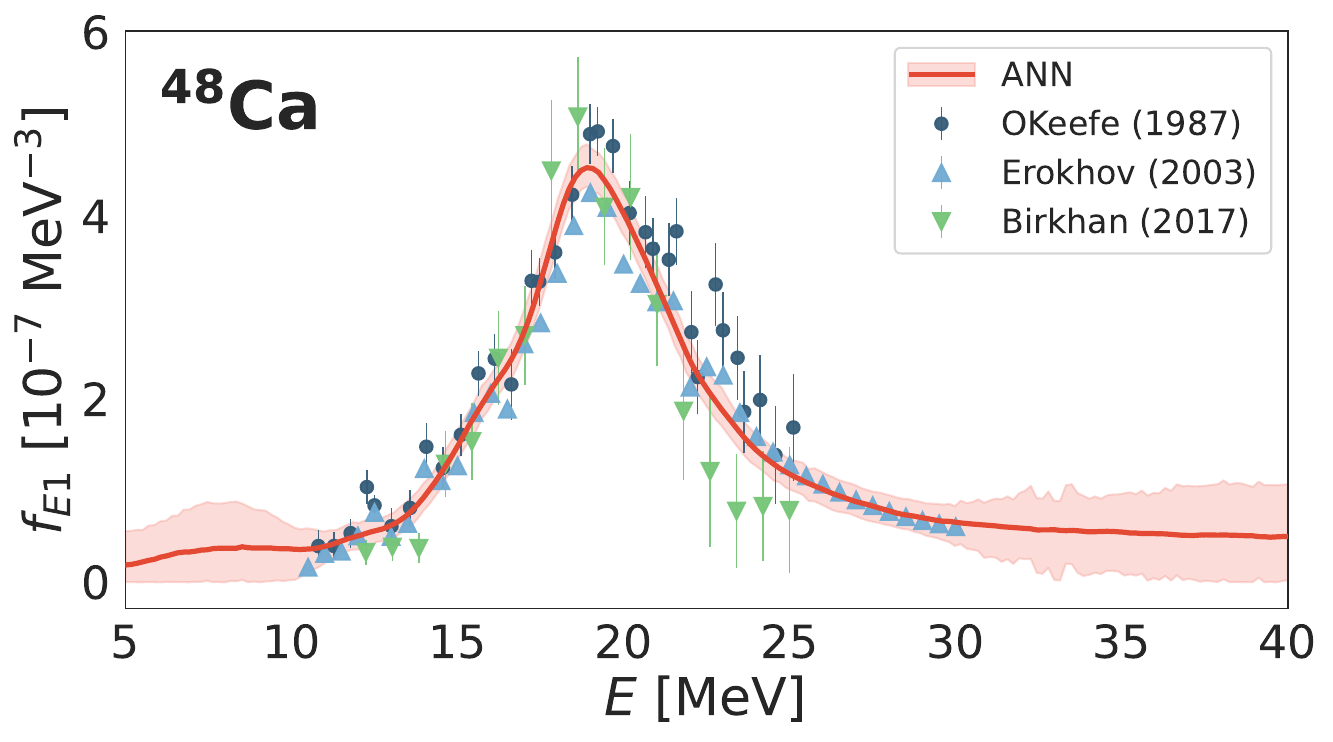}
\caption{Examples of training nuclei: ANN predictions of the dipole strength function for $^{40}$Ca (upper panel) and $^{48}$Ca (lower panel) in comparison to experimental data from Refs.~\cite{dolbilkin1966,erokhova2003,fea23,ahrens1985,okeefe1987,Birkhan2017} (see text for details). %(Color online) Blue-toned symbols are used for photoabsorption/photoneutron data, while the green-tones symbols denote the $(p,p')$ data.
}
\label{FIG1}
\end{figure}

\begin{figure}[h]
\includegraphics[width=1 \linewidth]{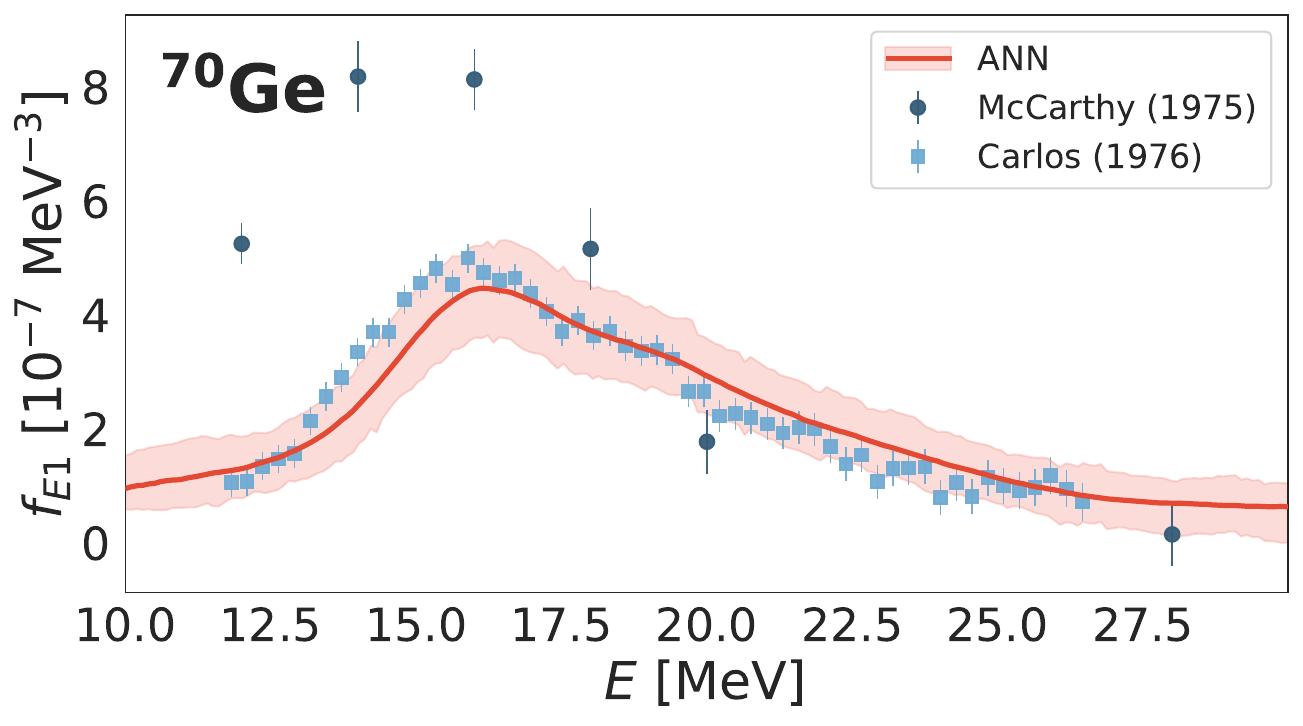}
\includegraphics[width=1 \linewidth]{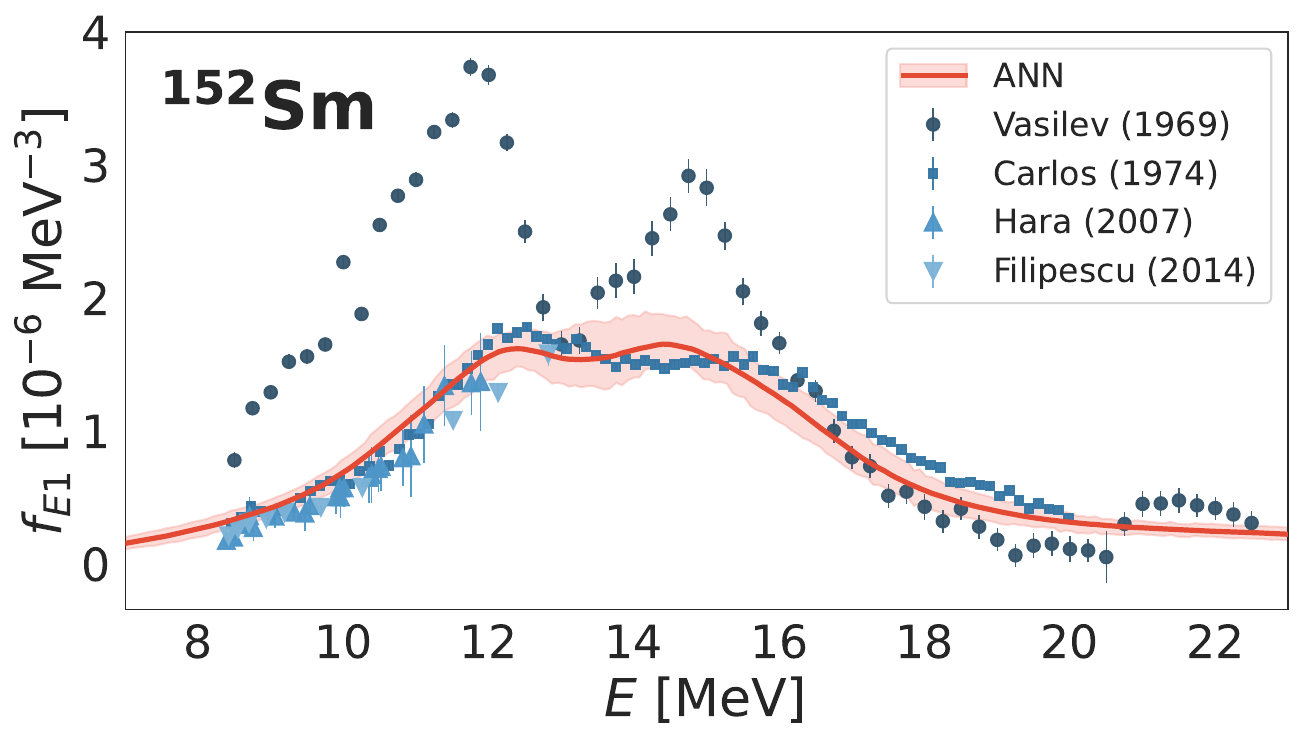}
\caption{Examples of testing nuclei: ANN predictions of the  dipole strength function for $^{70}$Ge (upper panel) and $^{152}$Sm (lower panel), in comparison to photo-nuclear experimental data from Refs.~\cite{mccarthy1975,carlos1976,vasilev1969,carlos1974,hara2007,filipescu2014} (see text for details).
}
\label{FIG2}
\end{figure}
We begin discussing our results by first displaying two nuclei used in the training phase, namely $^{40}$Ca and $^{48}$Ca, shown in  Figure~\ref{FIG1}.
 For $^{40}$Ca (upper panel), three experimental datasets from photoabsorption exist by Dolbilkin et al.~\cite{dolbilkin1966}, Ahrens et al.~\cite{ahrens1985}, and Erokhova et al.~\cite{erokhova2003}, plus one  $(p,p')$ dataset from Fearick et al.~\cite{fea23}.
 For $^{48}$Ca (lower panel), two photoabsorption datasets are available by Erokhova et al.~\cite{erokhova2003} and O'Keefe et al.~\cite{okeefe1987}, with an additional $(p,p')$ dataset by Birkhan et al.~\cite{bir17}. For both nuclei, the experimental data provide a relatively consistent picture, in the sense that data agree within error bars, or when data have no error bars they follow the trend, except in the regions $10 - 12$ MeV and $23 - 27$ MeV for $^{40}$Ca. All of these data (but no error bars) are seen by our ANN in the training phase.

After the training, the ANN predicts an $f_{E1}$ which is about the average of the experiments in the energy region covered by more than one experiment, albeit presents a lower uncertainty than that of the available ones from the experiment themselves, because new information on other nuclei is learned by the network. In particular for $^{48}$Ca below 10 MeV and above 30 MeV the uncertainty of the ANN increases due to the fact that no data exist and therefore this region is less constrained.
A similar pattern is found for other training nuclei.

At this point, it is interesting to  take a look at the ANN predictions for the 10 nuclei that were excluded from the training, namely
 $^{19}$F, $^{58}$Ni, $^{70}$Ge, $^{89}$Y, $^{114}$Sn,  $^{146}$Nd, $^{152}$Sm, $^{174}$Yb,
 $^{207}$Pb, and $^{234}$U. 
The ensemble of excluded nuclei was generated through a combination of random selection and deliberate intent. In particular, we kept nuclei outside of the training set when their corresponding data were drastically in conflict with each other. 

In Figure~\ref{FIG2}, we present results for two selected test nuclei, namely $^{70}$Ge and $^{152}$Sm, where something interesting can be learned. 
For $^{70}$Ge (upper panel), the photoabsorption data by McCarthy et al.~\cite{mccarthy1975} are much higher than the photoneutron data of Carlos et al.~\cite{carlos1976}. 
For $^{152}$Sm (lower panel), four photoabsorption datasets exist:
the older set of data by Vasilev et al.~\cite{vasilev1969}
presents a clearly higher structure with two bumps peaked at about 11 and 15 MeV, while the other datasets either cover only low energy like those by Fillipescu et al.~\cite{filipescu2014} and Hara et al.~\cite{hara2007}, or show primarily one big bump with perhaps a bit of a two-shoulder structure (the set by Carlos et al.~\cite{carlos1974}).

Interestingly, the ANN predictions with the corresponding uncertainties do not favor the data of \cite{mccarthy1975} and \cite{vasilev1969} for $^{70}$Ge and $^{152}$Sm, respectively, suggesting they do not follow the trend learned in the training set. 
Therefore, one could conclude that such dataset could be discarded. 
The large deviation of these two results from the other data may be related to the experimental technique based on unfolding bremsstrahlung excitation functions with the Penfold-Leiss method \cite{penfold59}, which is known to produce extremely large systematic uncertainties when the bremsstrahlung spectra are known with insufficient precision.

\begin{figure}[h]
\includegraphics[width=1 \linewidth]{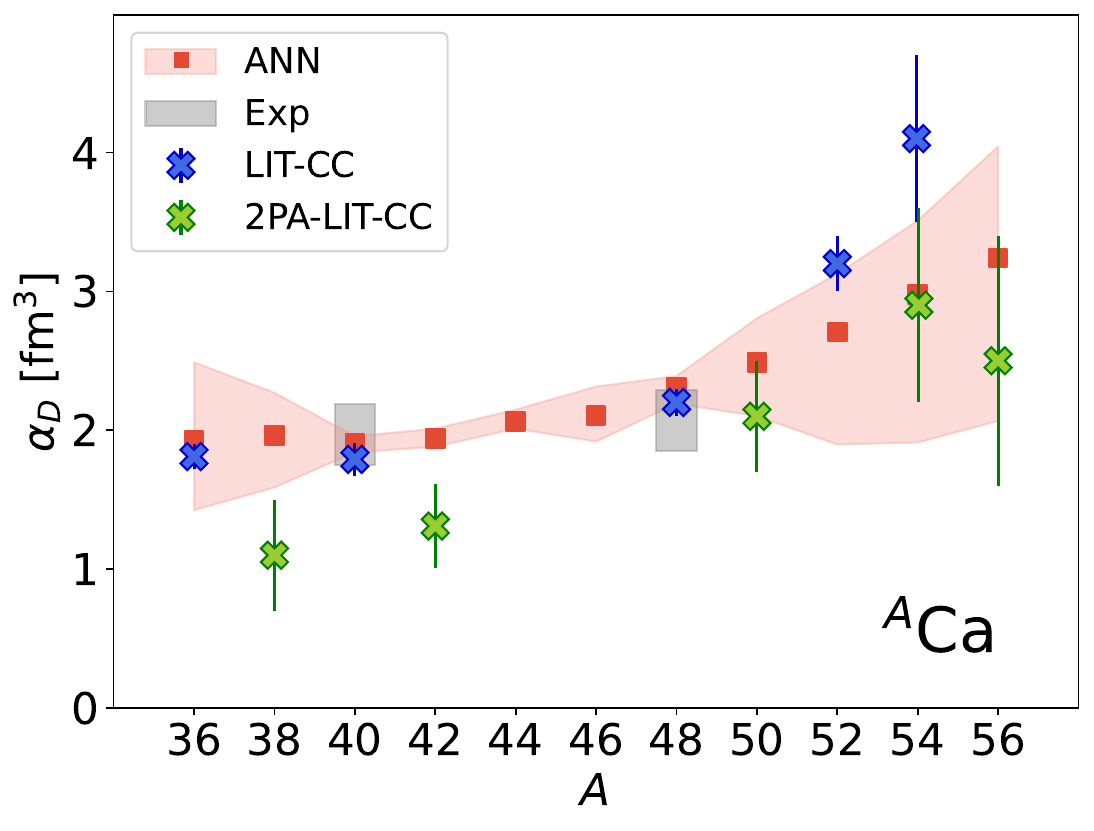}
\caption{Comparison to many-body theory:  ANN $\alpha_D$ predictions for the calcium isotope chain as a function of mass number $A$ in comparison to coupled-cluster calculations and experimental data from Ref.~\cite{fea23,Birkhan2017} (see text for details). Note that we integrate the predicted ANN strength functions from 10 to 40 MeV for consistency with the two experiments. }
\label{FIG3}
\end{figure}

Having established the predictive power of our  ANN on the dipole strength, we can now use these data to compute the  electric dipole polarizability $\alpha_D$  from Eq.~(\ref{alphaD}) and compare them for example against 
{\it ab initio} many-body calculations.
We focus on the calcium isotope chain, for which coupled-cluster results have recently been provided. Calculations for closed-shell nuclei with the Lorentz integral transform coupled-cluster (LIT-CC) method were provided in Refs.~\cite{hag16,miorelli2016,miorelli2018,fea23}, which were then extended in Ref.~\cite{bonaiti2024} to nuclei with two nucleons outside of a closed (sub)shell core via the two-particle attached formulation  (2PA-LIT-CC).
Figure~\ref{FIG3}  shows the ANN predictions, indicated by the (red) squares with the corresponding asymmetric error bars represented by the (red-toned) shaded area, in comparison to coupled-cluster results and their related error bars. 

It is important to remark that the only experimental data that the ANN sees for the calcium isotopes are for $^{40}$Ca and $^{48}$Ca, as already shown in Fig.~\ref{FIG1}. The two gray boxes of Fig.~\ref{FIG3} correspond to the experimental evaluation of $\alpha_D$ from Refs.~\cite{Birkhan2017,fea23}. We observe that ANN predictions between these two nuclei are very precise, while for the mass number lower than 40 and mass number higher than 48 the error bar strongly increases. This indicates the power of the ANN in interpolating nuclei, while at the same time emphasizing its limitations in extrapolating.

It is interesting to notice that the ANN prediction agrees nicely with the LIT-CC result for the 
closed-(sub)shell 
nucleus $^{36}$Ca, reported in Ref.~\cite{bonaiti2024}. For $^{42}$Ca, instead, we see that the ANN prediction is larger than the 2PA-LIT-CC calculation.
This confirms our intuition, already reported in Refs.~\cite{bonaiti2024,brandherm2024a},
 that the 2PA-LIT-CC framework at the present three-particle one-hole level might not contain enough correlations to capture the correct physics of $\alpha_D$. Regarding the neutron-rich sector beyond mass number 48, 
  we see that the ANN seems to favour our
  2PA-LIT-CC. However, we notice that the extent of the error bar depends on the details of the training data. For example, excluding the Oslo data from the training set would lead to an enhancement of the ANN error bar beyond $^{48}$Ca, which would make the prediction compatible also with the LIT-CC closed shell result.
  Therefore, we would like to stress that new experimental measurements in the neutron-rich region are necessary not only to better constrain our ANN in the neutron-rich sector, but also to have a solid comparison with {\it ab initio} theory.

\begin{figure}[t]
\includegraphics[width=0.95\linewidth]{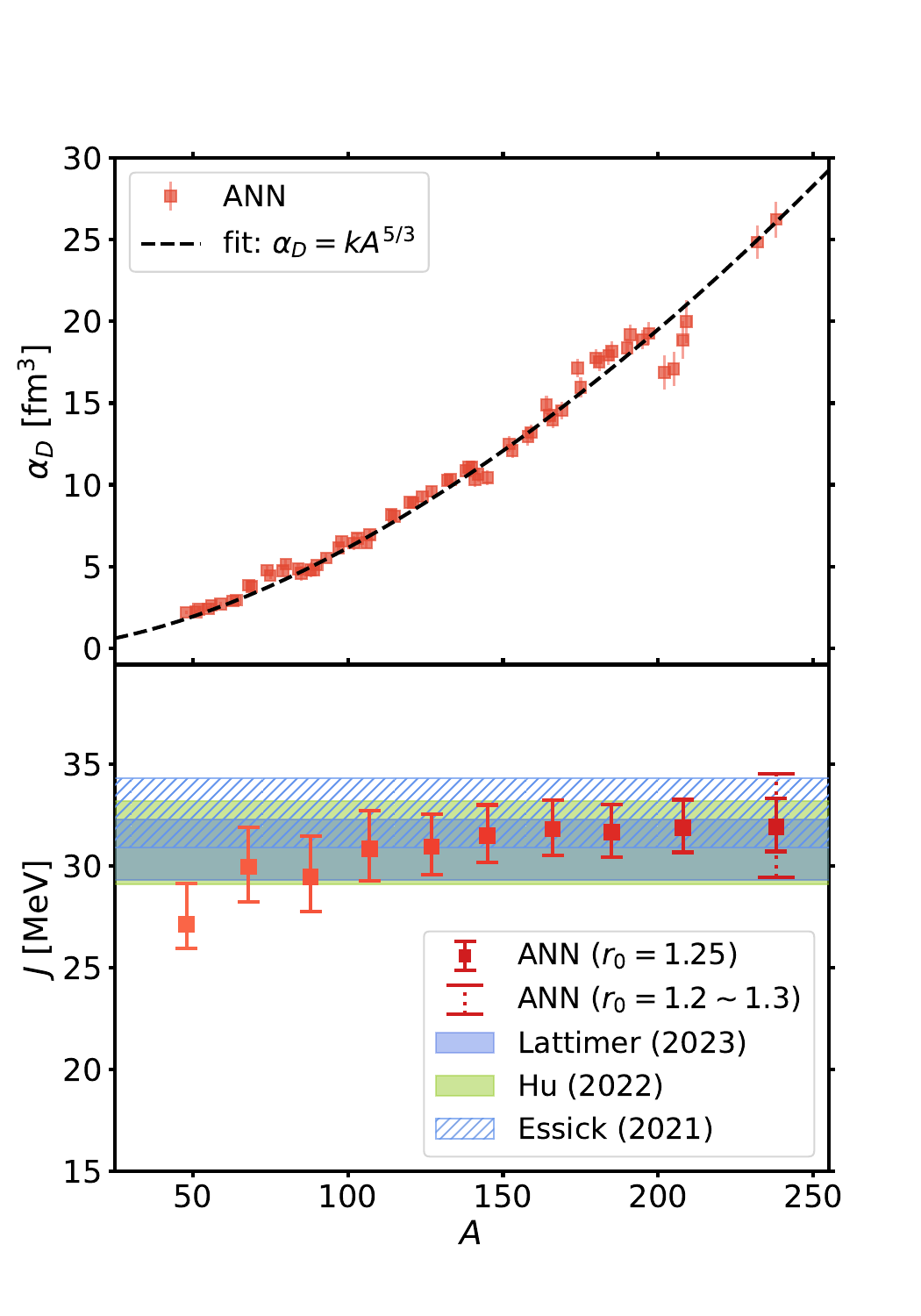}
\caption{(Upper panel) Predictions of $\alpha_D$, for isotopes with mass numbers $A$ from 48 to 238 obtained using our ANN. The dashed line represents a fit to 
Eq.~(\ref{migdal})
%a Migdal model formula \cite{migdal1945}, 
assuming $r_0 = 1.25$ fm. (Lower panel) Symmetry energy $J$ as a function of the maximum mass number $A$ included in the fit to the Migdal model, compared with available constraints in the literature \cite{lattimer2023,hu2022,essick2021}. The solid red bars indicate the uncertainty in the fitted $J$ arising from errors in the ANN predictions, while the dotted red bars represent the uncertainty in $J$ associated with variations in the $r_0$ parameter within the range of 1.2 fm to 1.3 fm.}
\label{FIG4}
\end{figure}

Finally, we address the connection between the dipole polarizability and the EoS of neutron-rich matter, which arises from the polarizability's sensitivity to the symmetry energy and its density dependence. The symmetry energy governs how the energy of nuclear matter changes as the neutron-proton ratio is varied, which in turn affects the properties of neutron-rich systems, including neutron stars. Precise knowledge of $\alpha_D$ can help refine the EoS, improving our understanding of neutron star properties. 
We will focus on a possible strategy to extract information on the symmetry energy based on our ANN predictions for $\alpha_D$. Thanks to the ANN, in fact, we can provide predictions of this observable from light to heavy-mass nuclei. This opens up the opportunity of employing phenomenological formulas describing the scaling of $\alpha_D$ with mass number to extract information on the nuclear EoS. 

The simplest way of modeling the behavior of the polarizability as a function of mass number is given by the Migdal model~\cite{migdal1945,meyer1982}. In this approach, a hydrodynamic model of the nucleus is adopted made by compenetrating neutron and protons fluids, which leads to
\begin{equation}
    \alpha_D = 2\alpha \hbar c \frac{A r^2}{48J} = 2\alpha \hbar c \frac{r_0^2 A^{5/3}}{48J}\,,
\label{migdal}
\end{equation}
where $\alpha$ is the fine-structure constant, $r$ is the nuclear radius parameterized as $r = r_0 A^{1/3}$, and $J$ is the symmetry energy at saturation density entering the EoS.
We now fit this formula to our ANN prediction in order to extract an estimate of $J$. Predictions for the polarizability are obtained by integrating the ANN dipole strength function from 2 to 40 MeV, where the integral has converged. For each isotopic chain, we include in the fit the most abundant stable isotope. In this region, in fact, we can rely on the good interpolation capabilities of our ANN where no data are available. Given that  light nuclei may deviate from the Migdal approach because of surface contributions to the symmetry energy~\cite{vonneumanncosel16},  we restrict our fit to $A\ge 48$. 

Our results are summarized in Fig.~\ref{FIG4}. In the upper panel, we show our ANN predictions for the polarizability as a function of mass number, accompanied by a fit to Eq.~(\ref{migdal}), where we assume a characteristic value of $r_0 = 1.25$ fm \cite{krane1991}. Despite its approximate nature, the fit nicely encompasses our ANN predictions for $\alpha_D$. We observe significant deviations only in the heavy-mass region, around $A = 200$. We point out that this effect may arise from a large number of measurements for $^{205}$Tl, $^{208}$Pb, and $^{209}$Bi yielding a lower dipole strength, while a smaller subset (including the most recent experiment) gives higher values. This distribution leads to a reduced ANN prediction for $\alpha_D$, which deviates from the Migdal prediction. 
In the lower panel, we illustrate how the extracted value of $J$ changes by gradually increasing the maximum mass number $A$ of the nuclei included in the fit up to $238$.

The extraction of $J$ is affected by two sources of uncertainty: the error of our ANN predictions of $\alpha_D$, and our knowledge of the $r_0$ parameter entering the empirical formula for the nuclear radius. To propagate the uncertainty of the ANN results for $J$, we evaluate the spread in $J$ by fitting the Migdal model to the upper and lower bounds of the ANN predictions. Such procedure leads to the solid (red) error bars in the figure. As for the uncertainty related to $r_0$, shown by the dotted (red) bar, we take the variation between the results obtained with $r_0 = 1.2$ fm and $r_0 = 1.3$ fm as our error estimate. It is worth pointing out that the error related to the choice of $r_0$, intrinsic to the Migdal model, is dominating, amounting to almost twice the ANN uncertainty.

Considering up to $A=238$ in our fit, we obtain $J = 32(3)$ MeV. It is interesting to observe that our ANN-based result for $J$ saturates already when including in the fit all nuclei with $A<150$, indicating that valuable information on nuclear matter can be obtained also from this lower-mass region. Our result is compared to different values from the literature in Fig.~\ref{FIG4} (lower panel). The value by Lattimer~\cite{lattimer2023} relies on a global analysis of experimental data of neutron skin thicknesses in $^{48}$Ca and $^{208}$Pb. The result by Hu et al.~\cite{hu2022} is extracted from nuclear matter calculations using non-implausible chiral effective field theory interactions, while that from Essick et al.~\cite{essick2021} considers a Gaussian process model of the EOS constrained by chiral forces and astrophysical observations. Our estimate for $J$ is in excellent agreement with the above-mentioned different approaches, corroborating once more the robustness of our trained ANN.

%\twocolumn
%%%%%%%%%%%%%%%%%%%%%%%%%%%%%%%%%%%%%%%%%%%%%%%%%%%%%%%
% Conclusions
%%%%%%%%%%%%%%%%%%%%%%%%%%%%%%%%%%%%%%%%%%%%%%%%%%%%%%%
\section{Conclusions}

In this work, we performed the first extraction of dipole strength functions using machine learning techniques for several nuclei across the nuclear chart.
We observed that the ANN effectively learns information about various nuclei, leading to predictions that align well with the data for training and validation nuclei, as well as most test nuclei. When faced with conflicting data, the ANN tends to favor one trend over others. In some cases, specific datasets disfavored by the ANN, e.g., those from Ref.~\cite{mccarthy1975} for $^{70}$Ge and those from Ref.~\cite{vasilev1969} for $^{152}$Sm,  could also be discarded based on experimental considerations. This reinforces the ANN ability to identify potentially unreliable data.

We confronted the ANN predictions also with recent {\it ab initio} coupled-cluster calculations of the electric dipole polarizability in the calcium isotope chain. We found that generally up to $^{50}$Ca, the ANN prediction agrees well with calculations of closed-(sub)shell nuclei, but disagrees with calculations on open-shell nuclei. We argued that this might actually be due to a shortcoming of the many-body method. We observed that ANN uncertainties in the neutron-rich region beyond  $^{50}$Ca become large. Additional experimental data are essential to further refine and constrain the ANN in the neutron-rich sector, enhancing its accuracy and reliability.

Finally, assuming the simple Migdal model we extracted a value of the symmetry energy using the ANN predictions for the electric dipole polarizability. We found results that are in agreement with other recent determinations based on theory or a combination of theory and experiment. We observe that the uncertainty related to the ANN error is smaller with respect to the uncertainty intrinsic to the parameters of the Migdal model itself.
The interesting question of whether constraining the EoS using stable nuclei differs from using unstable nuclei cannot be fully addressed until more experimental data from the neutron-rich sector are obtained.

In summary, our findings demonstrate that machine learning tools, particularly artificial neural networks, excel at interpolating among known nuclei but show limitations when extrapolating beyond the training data. This emphasizes the importance of expanding the available experimental datasets, especially in the neutron-rich sector,  and at the same time highlights the potential of artificial intelligence to complement traditional methods in nuclear research and guide future experimental efforts in the quest for a comprehensive understanding of nuclear properties.

\section*{Acknowledgments}
We acknowledge useful discussions with J.E.~Sobczyk, F. Marino and L.~Doria. This work was supported by the Deutsche Forschungsgemeinschaft (DFG)
through the Cluster of Excellence ``Precision Physics, Fundamental
Interactions, and Structure of Matter" (PRISMA$^+$ EXC 2118/1) funded by the
DFG within the German Excellence Strategy (Project ID 390831469), 
and through Project-ID 279384907 - SFB 1245, and by the U.S. Department of Energy, Office of Science,
Office of Nuclear Physics, under the FRIB Theory Alliance award DE-SC0013617, and Office of Advanced Scientific Computing Research and Office of Nuclear Physics,
Scientific Discovery through Advanced Computing (Sci-DAC) program (SciDAC-5 NUCLEI). Oak Ridge National Laboratory is
supported by the Office of Science of the Department of
Energy under contract No. DE-AC05-00OR22725.  Computations were performed on the supercomputer MogonII at Johannes Gutenberg-Universit\"{a}t Mainz.

%%%%%%%%%%%%%%%%%%%%%%%%%%%%%%%%%%%%%%%%%%%%%%%%%%%%%%%
% Bibliography
%%%%%%%%%%%%%%%%%%%%%%%%%%%%%%%%%%%%%%%%%%%%%%%%%%%%%%%

\bibliography{biblio.bib}

\end{document}

\begin{frontmatter}

\title{Supplementary Material}

\author[kph]{Weiguang Jiang}
\ead{wjiang@uni-mainz.de}
\author[kph]{Tim Egert}
\ead{tiegert@students.uni-mainz.de}
\author[kph,him]{Sonia Bacca}
\ead{s.bacca@uni-mainz.de}
\author[kph,frib,ornl]{Francesca Bonaiti}
\ead{bonaiti@frib.msu.edu}
\author[da]{Peter von Neumann Cosel}
\ead{vnc@ikp.tu-darmstadt.de}
\address[kph]{Institut f\"{u}r Kernphysik and PRISMA$^+$ Cluster of Excellence, Johannes Gutenberg-Universit\"at, 55128 Mainz, Germany}
\address[him]{Helmholtz-Institut Mainz, Johannes Gutenberg-Universit\"at Mainz, D-55099 Mainz, Germany}
\address[frib]{Facility for Rare Isotope Beams, Michigan State University, East Lansing, MI 48824, USA}
\address[ornl]{Physics Division, Oak Ridge National Laboratory, Oak Ridge, TN 37831, USA}
\address[da]{Institut f\"{u}r Kernphysik, Technische Universität Darmstadt, D-64289 Darmstadt, Germany}

\renewenvironment{abstract}{}{}
\end{frontmatter}

\section{Artificial Neural Network (ANN) architecture} 
\begin{figure}[bht]
\includegraphics[width=1 \linewidth]{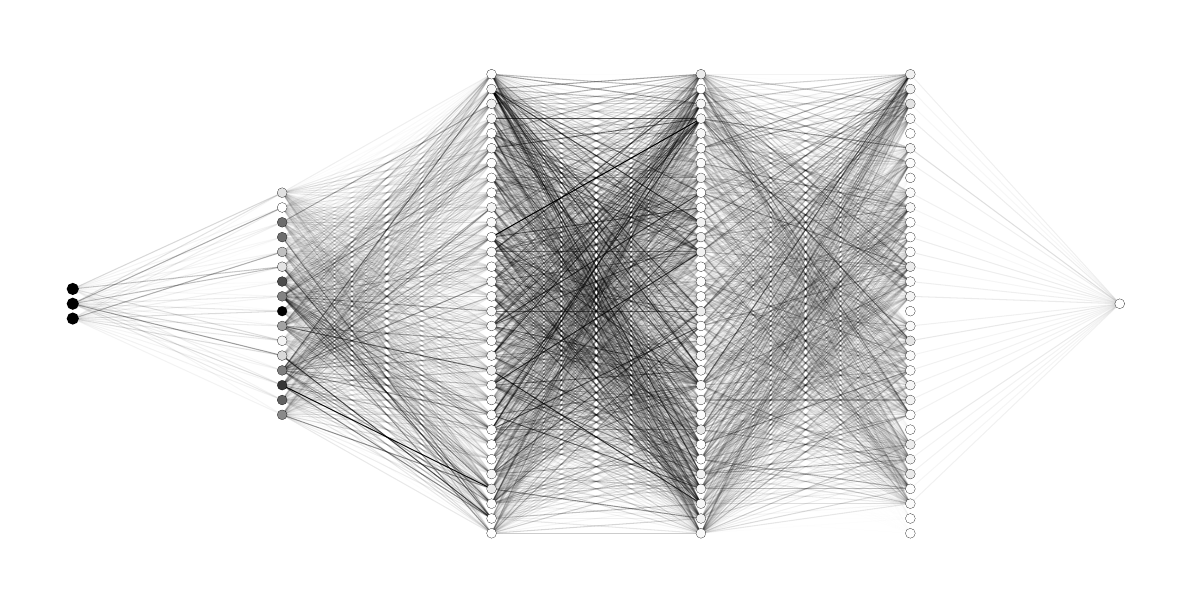}
\caption{Schematic representation of one of the neural networks used in this study. The figure illustrates the absolute values of the weights (lines) and biases (dots), depicted in grayscale and normalized to the range [0,1], where white corresponds to zero and black represents one.}
\label{NN_architecture}
\end{figure}

The artificial neural network employed in this study is a fully connected feedforward network designed to predict the dipole strength function $f_{E1}$ across the nuclear chart. Figure \ref{NN_architecture} shows the schematic structure of the network. The input layer includes the mass number $A$, proton number $Z$, and gamma-ray energy $E$, while the output layer predicts the preprocessed $f_{E1}$. Between these layers, the network comprises four hidden layers with [16, 32, 32, 32] neurons. Non-linear relationships are captured through tanh activation functions, allowing the network to learn complex patterns in the data while maintaining smooth and continuous behavior. Note that for the output layer, the activation function is chosen to be ReLU to ensure a physical non-negative $f_{E1}$ prediction.

The RMSE is used as the evaluation metric, offering a scale-aligned and interpretable measure of prediction performance. The ANN weights are initialized using Glorot (Xavier) initialization, which draws samples from a uniform distribution within
\begin{equation}
    [\ -l\ ,\ l\ ], \ \text{where} \ l = \sqrt{\frac{6} { n_{\text{in}} + n_{\text{out}}}},
\end{equation}
with $n_{\text{in}}$ and $n_{\text{out}}$ representing the number of input and output neurons in each layer.  This initialization maintains the variance of activations across layers, aiding smooth convergence.

The current architecture was selected after iterative testing to balance model complexity and overfitting risks. Given the current input data size (166,370 data points after augmentation) and the number of trainable parameters (2,753), there is still potential to further deepen or widen the network before encountering significant overfitting. This suggests that the network's capacity could be increased if more intricate patterns in other new data need to be captured.

\section{Training Process}
\begin{figure}[bht]\centering
\includegraphics[width=0.7 \linewidth]{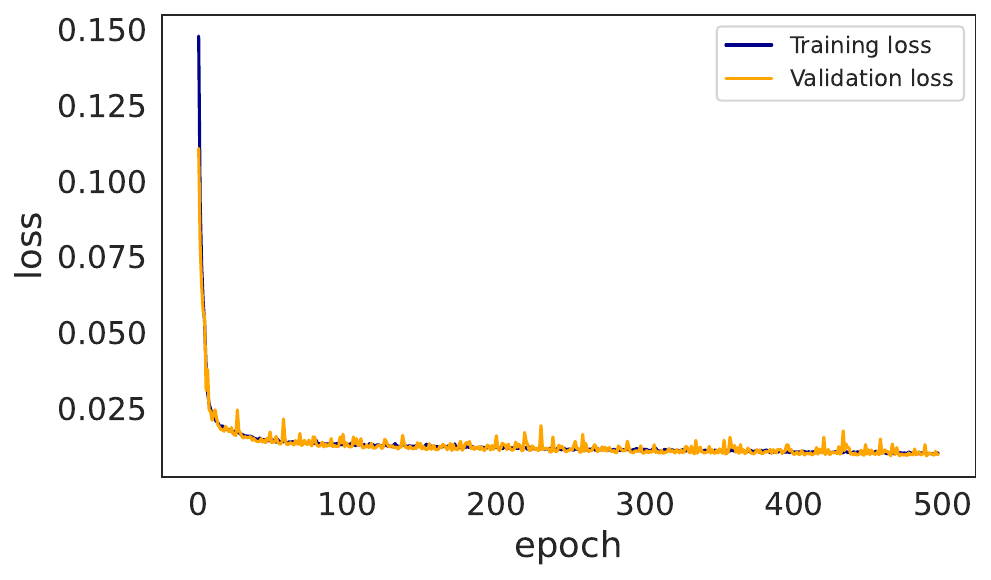}
\caption{Training and validation loss as a function of epochs, illustrating the convergence behavior of the network during training.}
\label{Losses}
\end{figure}

The training process uses the Adam optimizer, configured with an initial learning rate of $\eta = 0.001$. Adam combines the benefits of the Adaptive Gradient Algorithm (AdaGrad) and the Root Mean Square Propagation (RMSProp), ensuring efficient optimization even for large datasets. Early stopping with patience of 50 epochs is employed to prevent overfitting, monitoring validation loss to halt training when improvement plateaus. Figure~\ref{Losses} shows a typical learning loss curve for our ANN model. As we can see the network is already getting well-trained after $\sim$~100 epochs and the performance of validation data is as good as the training ones indicating the possible overfitting is under control. The small fluctuations in the validation loss, while the neural network is performing well, typically occur in well-trained models and are usually due to the inherent randomness of the optimization process and minor changes in gradient estimates. 

During training, we observed that proper alignment between the model’s hyperparameters, such as the initial weights and starting learning rate, and the scale of the input data is crucial for effective optimization. This alignment ensures that gradients computed during backpropagation remain within a suitable range, enabling the optimization process to proceed efficiently and accurately. Rather than directly adjusting hyperparameters like the learning rate or weight initialization, we achieve an equivalent effect by rescaling the input data to align with these parameters. Specifically, the dipole strength data, which serves as the input, are scaled by a factor of $10^8$ to align with the starting learning rate $\eta = 0.001$. This scaling step ensures that the input data and weight updates operate on compatible magnitudes, preventing issues like vanishing or exploding gradients during training. Additionally, we address variability in the scale of dipole strength values across different nuclei by normalizing these values for each nucleus. This is approximately achieved by dividing the dipole strength by the corresponding mass number $A$, providing a rough normalization to ensure general consistency across the dataset.

These preprocessing steps effectively standardize the input data, allowing the network to learn generalizable patterns across the entire nuclear chart. By rescaling the inputs rather than fine-tuning the hyperparameters themselves, we achieve a balance between efficient training and robust model performance while preserving the interpretability of the input features.

\begin{figure}[bht]\centering
\includegraphics[width=0.7 \linewidth]{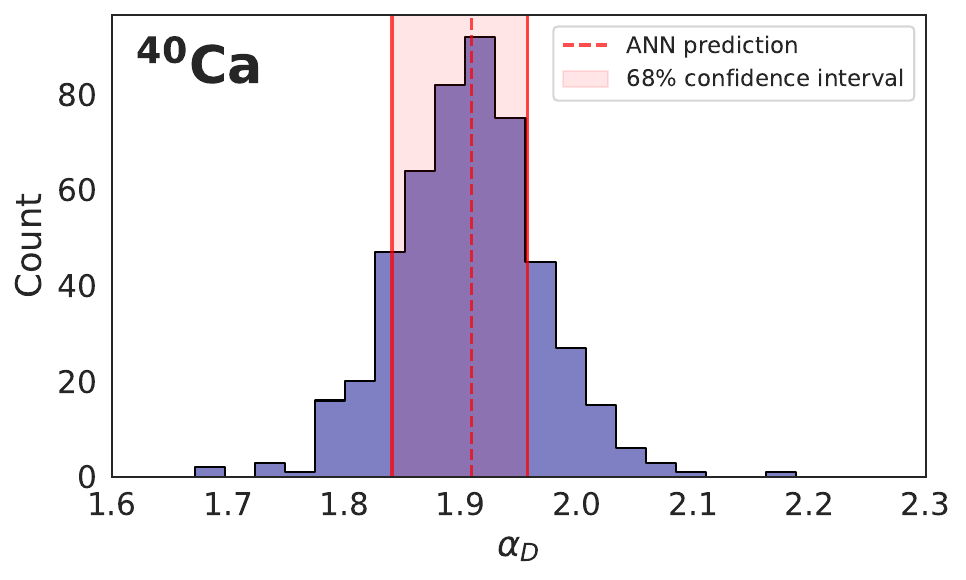}
\caption{ Predicted distribution of $\alpha_D$ for $^{40}\text{Ca}$ generated using an ensemble of 500 ANNs.}
\label{Ensamble_model}
\end{figure}
\section{Uncertainty Quantification}
To quantify uncertainties and improve robustness, we train an ensemble of 500 ANNs with different weight initializations and randomized splits of the data between training and validation sets. This ensemble approach serves two purposes. First, for uncertainty quantification, it captures variability in predictions due to parameter uncertainty and partially addresses data uncertainty from variability in the training data. Second for robustness, it balances the effects of beneficial and detrimental random initializations. Figure~\ref{Ensamble_model} shows the histogram of ensemble predictions of $^{40}$Ca $\alpha_D$ using 500 ANNs. As we can see, the target observable exhibits an unimodal distribution, which is advantageous for our uncertainty analysis. This allows us to use the median (or mean) of the distribution as the network prediction, while the 68\% highest density interval serves as the uncertainty region, providing a reasonable measure of variability around the central value.